\documentclass[10pt,aps,nofootinbib,superscriptaddress,twocolumn,preprintnumbers,balancelastpage]{revtex4-1}

\usepackage{graphicx}
\usepackage{epstopdf}
\usepackage{bm}

\usepackage[colorlinks]{hyperref}
\hypersetup{
     colorlinks   = true,
     citecolor    = red,
 linkcolor=blue
}
\usepackage{color}
\usepackage{soul}
\usepackage{amsmath}
\usepackage{amssymb}
\usepackage{mathtools}
\usepackage{xfrac}

\newcommand{\Pb}{{\rm Pb}}
\newcommand{\pp}{{{\rm p}-{\rm p}}}
\newcommand{\gaga}{{\gamma\gamma}}

\begin{document}

\newcommand{\toname}{{\sqrt{s^{c}_{\gaga}}}}

\newcommand{\ion}{{\textrm{ion}}}
\newcommand{\Nu}{{\textrm{nu}}}
\newcommand{\GeV}{{\rm GeV}}
\newcommand{\TeV}{{\rm TeV}}
\newcommand{\fm}{{\rm fm}}
\renewcommand{\max}{{\rm max}}

\graphicspath{{Figures/}}

\title{
Searching for axion-like particles with ultra-peripheral heavy-ion collisions
}

\author{Simon Knapen}
\author{Tongyan Lin}
\author{Hou Keong Lou}
\author{Tom Melia}
\affiliation{Department of Physics, University of California, Berkeley, California 94720, USA}
\affiliation{Theoretical Physics Group, Lawrence Berkeley National Laboratory, Berkeley, California 94720, USA}

\begin{abstract}
We show that ultra-peripheral  heavy-ion collisions at the LHC can be used to search for axion-like particles with mass below 100 GeV. 
The $Z^4$ enhanced photon-photon luminosity from the ions provides a large exclusive production rate, with a signature of a resonant pair of back-to-back photons and no other activity in the detector. 
In addition, we present both new and updated limits from recasting multi-photon searches at LEP II and the LHC, which are more stringent than those currently in the literature for the mass range 100 MeV to 100 GeV.
\end{abstract}

\date{\today}

\maketitle

\section{Introduction}
\label{sec:intro}

A number of outstanding experimental and theoretical observations point to an incompleteness of the standard model (SM); notable examples include the existence of dark matter, the strong CP problem, and the hierarchy problem.  Proposed resolutions typically involve the introduction of new particles or even whole new sectors beyond the SM. 
The Large Hadron Collider (LHC), in its capacity as a energy-frontier proton-proton ({\rm p}-{\rm p}) collider, has a suite of dedicated searches for many different new physics scenarios (for an overview, see Ref.~\cite{ATLAS:1999vwa,Ball:2007zza}).

Beyond \pp{} collisions, the LHC also collides heavy ions at unprecedented energies. ATLAS, CMS, LHCb and ALICE have all recorded proton-lead ({\rm p}-{\rm Pb}) and lead-lead (\Pb{}-\Pb{}) collisions.
For Pb-Pb collisions at the LHC, the design luminosity is $\sim 1\, \textrm{nb}^{-1}/$year, with an eventual center-of-mass energy per nucleon of $\sqrt{s_{NN}}=5.5$ TeV. With this reduced luminosity and lower per-nucleon collision energy, heavy-ion collisions are not optimized for typical beyond the SM (BSM) physics searches. 
However, the large charge of the lead ions ($Z=82$) results in a huge $Z^4$ enhancement for the coherent photon-photon luminosity, which can in principle be exploited to search for new physics that couples predominantly to photons.  Interestingly, this coherent enhancement extends to energies   above $100$ GeV, essentially because the wavelength of such high energy photons is still longer than the Lorentz-contracted size of the ultra-relativistic Pb ions.

These coherent electromagnetic interactions occur in ultra-peripheral collisions (UPCs), where the impact parameter is much larger than the ion radius, such that the ions scatter quasi-elastically and remain intact. (See Ref.~\cite{Baur:2001jj, Bertulani:2005ru, Baltz:2007kq} for reviews.) Such exclusive processes are characterized by a lack of additional detector activity and a large rapidity gap between the produced particles and outgoing beams. This allows  very efficient background rejection of non-exclusive interactions and provides a clean environment to search for new particles.  One particularly fascinating early proposal was a search for the SM Higgs boson in photon fusion \cite{0954-3899-15-3-001,Papageorgiu:1988yg,Drees:1989vq}. Although the rate for this process  is too small for the planned luminosity at the LHC~\cite{d'Enterria:2009er}, it is nevertheless a very instructive benchmark for the study of exclusive particle production in UPCs. Other  proposals include searches for {\it e.g.~}supersymmetry \cite{Greiner:1992fz} or extra dimensions \cite{Ahern:2000jn}, but have not been competitive with the analogous searches with p-p collisions.

\begin{figure}[b!]
\centering
\includegraphics[width=6cm]{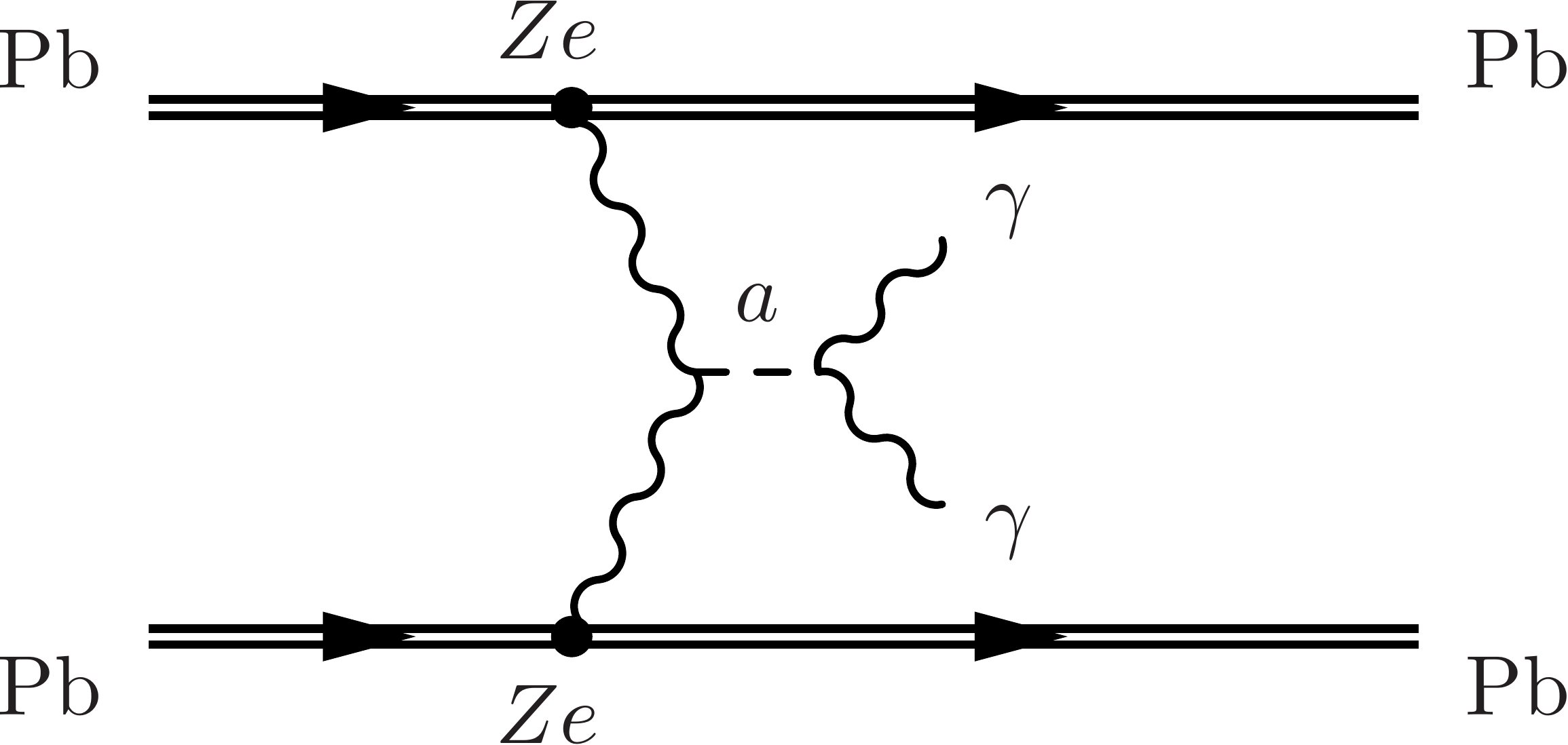}
\caption{Exclusive ALP production in ultra-peripheral Pb-Pb collisions.}
\label{fig:fey_alp}
\end{figure}

In this Letter, we present an application of heavy-ion collisions to search for scalar and pseudoscalar particles  produced in photon fusion (Fig.~\ref{fig:fey_alp}) and with mass in the range 5 to 100 GeV.  (See \cite{PhysRevD.34.2896,PhysRevLett.55.461,PhysRevLett.56.302}  for early proposals related to MeV-scale particles in low energy heavy ion collisions.) 
Relatively light pseudoscalar bosons are natural ingredients in a large class of models which invoke the breaking of approximate symmetries. The $\pi^0$ and $\eta$ are known examples in the SM.  
In extensions of the SM, such particles can couple to the electromagnetic sector through a Lagrangian of the form
\begin{align}\label{eq:lag}
  \mathcal{L}_a = 
  \frac{1}{2}(\partial a)^2 -\frac{1}{2}m_a^2 a^2- \frac{1}{4}\frac{a}{\Lambda} F \widetilde{F}\,,
\end{align}
where $a$ is the new pseudoscalar, often referred to as an axion-like particle (ALP), $\tilde F^{\mu\nu}\equiv\frac{1}{2}\epsilon^{\mu\nu\rho\sigma}F_{\rho\sigma}$, $m_a$ is the mass of the ALP, and ${1}/{\Lambda} $ is the coupling constant.  We also consider an ALP coupling to hypercharge, through the operator $-\frac{1}{4 \cos^2 \theta_W}\frac{a }{\Lambda} B \widetilde{B}$.  
Although we take a pseudoscalar as a benchmark, our conclusions apply for scalars as well, upon substituting $\tilde F$($\tilde B$) with $F$($B$) in Eq.~\eqref{eq:lag}. 
For UPCs, the total cross section for ALP production in the narrow width approximation is given by 
\begin{align}
  \sigma_{a} = \frac{8\pi^2}{m_a}\Gamma(a \rightarrow \gamma \gamma) \mathcal{L}_{\gamma\gamma}(m_a^2),
\end{align}
where $\Gamma(a \rightarrow \gamma \gamma)=\tfrac{1}{64\pi}\tfrac{m_a^3}{\Lambda^2}$ is the decay width of the ALP into photons, and $\mathcal{L}_{\gamma\gamma}(m_a^2)$ is the photon-photon luminosity, evaluated at $m_a$.


\section{The photon-photon luminosity \label{sec:alps}}

For an ultra-relativistic charged particle with charge $Z\gg1$, the surrounding electromagnetic fields can be thought of as a pulse of nearly on-shell photons. This is known as the Weizs\"acker-Williams method, or the equivalent photon approximation  \cite{vonWeizsacker:1934nji,Williams:1934ad}. 
To facilitate a qualitative discussion of the relevant physics, we first consider the flux per unit area for a relativistic point particle: 
\begin{equation}
N(E,\vec b)=\frac{Z^2\alpha }{\pi^2 }\left(\frac{E}{ \gamma}\right)^2K_1^2\left( \frac{E |\vec b|}{ \gamma} \right) \,,
\label{eq:flux_per_area}
\end{equation}
where $\gamma$ is the Lorentz boost of the ion, $\vec b$ is the transverse displacement from the moving charge, and $K_{1}$ is the modified Bessel function.
The total photon-photon luminosity for exclusive collisions is then  
 \begin{equation}\label{eq:masterform}
 \begin{split}
 \mathcal{L}_{\gamma\gamma}(\hat s)=\frac{1}{\hat{s}}\int\! d^2\vec b_1 d^2\vec b_2\, d E_1d E_2\; N(E_1,\vec b_1)N(E_2,\vec b_2)\\
 \times\, P(|\vec b_1 -\vec b_2|) \delta(\hat s-4 E_1 E_2)
 \end{split}
 \end{equation}
where $P$ is the probability for the absence of hadronic interactions. To further ensure that the collisions are exclusive, the integral is restricted to $|\vec b_{1,2}|>R_A$, where $R_A$ is the nuclear radius. This justifies the approximation of a point-like charge distribution in Eq.~\eqref{eq:flux_per_area}; different charge distributions have been considered as well, see {\it e.g.}~\cite{Baur:1991fn}.

If we neglect the exclusivity term $P$ 
Eq.~\eqref{eq:masterform} can be written in a factorized analytic form \cite{Baur:1990fx,Cahn:1990jk}
\begin{align}
\begin{split}
\mathcal{L}_{\gamma\gamma}(\hat s) =
\frac{1}{\hat{s}}  \int d E_1 d E_2\;
  n_{\gamma}\!\left(\frac{E_1}{E_{R}}\right)     n_{\gamma}\!\left(\frac{E_2}{E_{R}}\right)   \\
  \,\times\,  \delta(\hat s-4E_1 E_2 )\,,
\end{split}
\label{eq:production}
\end{align} 
where $E_R=\gamma/R_A$, and
\begin{equation}\hspace{-.05cm}\resizebox{.44 \textwidth}{!} 
{
  $n_\gamma(x)=
  \tfrac{2Z^2 \alpha}{\pi}
  \left(x K_0(x)K_1(x) - 
     \tfrac{x^2}{2}\!\left[
    K_1^2(x)-K_0^2(x)
    \right]
  \right),$
}
\label{eq:photon_flux}
\end{equation}
which acts as a photon distribution function with $x=E_{1,2}/E_{R}$.
The center-of-mass energy where the coherently enhanced photon-photon luminosity becomes exponentially suppressed is roughly $2E_{R}$. For the LHC, this scale is 
\begin{align}
2\, E_{R}=\frac{2\gamma}{R_A} \simeq 170 \;\GeV \bigg(\frac{\sqrt{s_{NN}}}{5.5\; \TeV}\bigg)\bigg(\frac{7\; \fm}{R_A}\bigg)\,,
\end{align}
where the boost factor $\gamma= 2932$ and ${R_A\simeq 1.2 \,A^{1/3}\, \fm}$, with $A=208$ 
for the isotope used at the LHC. 

In general, the exclusivity factor $P$ can however not be neglected, which implies that Eq.~\eqref{eq:masterform} cannot be factorized. To address this issue, we implemented ALP production in the public Monte Carlo code \texttt{STARlight}~\cite{starlight}, which computes $P$ using nuclear density profiles and evaluates  Eq.~\eqref{eq:masterform} numerically. To estimate uncertainties associated with the ion radius, we also numerically evaluate Eq.~\eqref{eq:masterform} with an approximate form for $P(|\vec b_1 -\vec b_2|)=\theta(|\vec b_1 -\vec b_2|-2R_A)$. A $5$-$10\%$ variation of $R_A$ in this calculation translates into a $10$-$20\%$ effect on the luminosity. While we use the full \texttt{STARlight} calculation for the ALP signal, it is useful to compare with the simplified analytic form in Eq.~\eqref{eq:production}. In particular, for $E\ll E_R$ the flux is dominated by the \mbox{$|\vec b_1 -\vec b_2|\gg2R_A$} part of the integral, and the effect of the exclusivity factor is subdominant. Compared to the analytic approximation in Eq.~\eqref{eq:production}, we find that
the production cross section from {\tt STARlight} does not differ significantly for low mass ALPs, and is up to $20\%$ lower for $m_a \sim 100$ GeV.

Before moving to the experimental side of our story, there are a few additional details worth mentioning. First, in our treatment we neglected polarization effects which a priori could result in different production cross sections between scalars and pseudoscalars. However it was shown such effects are small, after integrating over all impact parameters~\cite{Greiner:1990aq,Baur:1990fx}. Second, in the equivalent photon approximation we implicitly assumed zero virtuality for the photons. The {\tt STARlight} calculation accounts for small photon virtualities, which results in a small ($\lesssim 100$ MeV) recoil of the $\gamma\gamma$ system against the ions themselves, see \cite{Baltz:2009jk}. Third, in a large fraction of events the ion will end up in an excited state after the collision takes place, for instance through giant dipole resonances \cite{RevModPhys.47.713}.  This can lead to downstream dissociation and neutrons ejected in this process can be picked up by the Zero Degree Calorimeters, providing an additional tagging technique for UPCs, see {\it e.g.}~\cite{Khachatryan:2016qhq}.

\section{Analysis strategy}
\label{sec:anal}

\textbf{Event selection:}
Our proposed analysis closely mimics the CMS  search for exclusive $\gamma\gamma$ production in p-p collisions~\cite{Chatrchyan:2012tv},  as well as the analogous search for light-by-light scattering proposed in Ref.~\cite{d'Enterria:2013yra}. For Pb-Pb collisions we rely on a dedicated trigger for UPC events which requires two photons with $p_T>2$ GeV, as well as the absence of any activity in at least one of the forward calorimeters. We hereby assume 90\% trigger and reconstruction efficiency. Our off-line selection consists of a pseudo-rapidity cut of $|\eta|<2.5$ on each photon as well as a tight cut on the azimuthal opening angle of $|\Delta\phi -\pi|<0.04$ for back-to-back photons. These cuts are nearly fully efficient for high mass ALPs and yield a 70\% fiducial efficiency for the lowest masses that can be recorded by the trigger ($m_a\gtrsim 5$ GeV). In the following, we assume a mass resolution of 0.5 GeV; although the mass resolution is expected to be worse for $m_a \gtrsim 20$ GeV, this will not affect the sensitivity since backgrounds are negligible in this regime.

We assume a rejection of all events from non-exclusive processes through a strong veto on tracks and additional isolated activity in the calorimeters, {\it e.g.}~\cite{Khachatryan:2016qhq}. Given the absence of pile-up at Pb-Pb collisions, we do not expect such a veto to significantly degrade the signal efficiency (see  also \cite{Chatrchyan:2012tv}). 

\textbf{Backgrounds:}
There are two types of backgrounds important for the ALP search: irreducible SM photon production and experimental backgrounds which fake di-photon production. The irreducible background consists of exclusive photon production mechanisms which give rise to an approximately smoothly falling distribution in $m_{\gamma \gamma}$. The second background comes from photon fakes due to electrons.

\begin{figure}[t!]
\centering
\includegraphics[width=3.4cm]{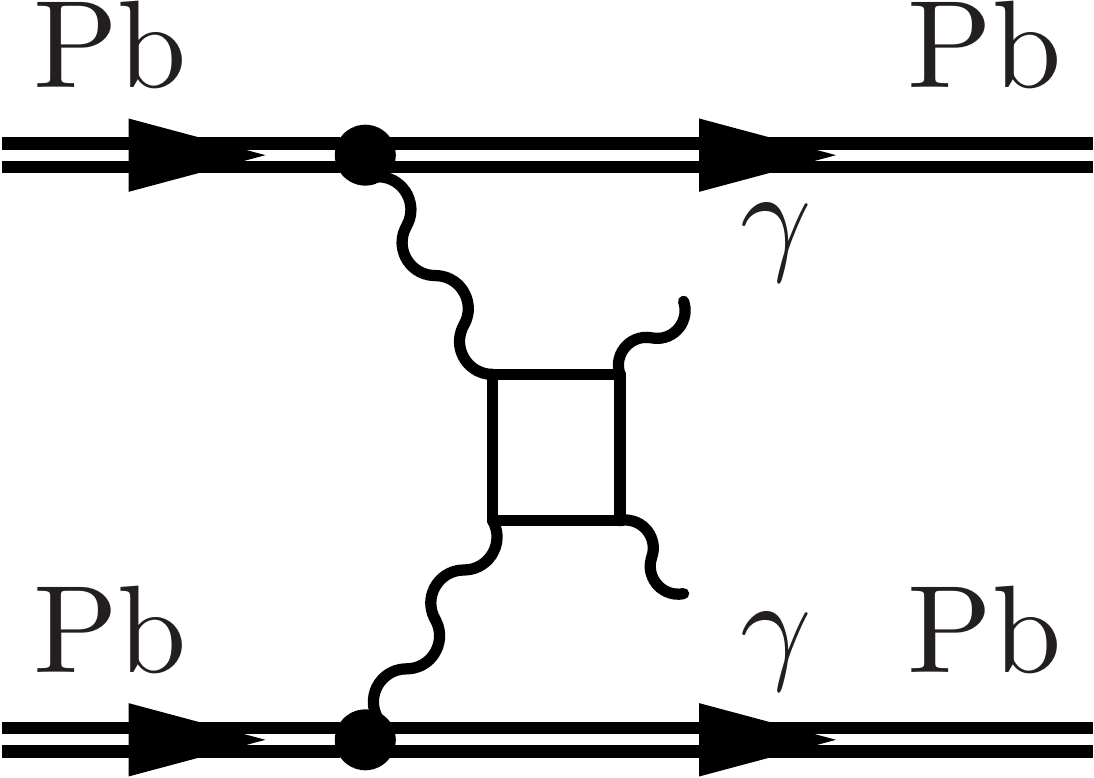}\qquad\qquad
\includegraphics[width=3.4cm]{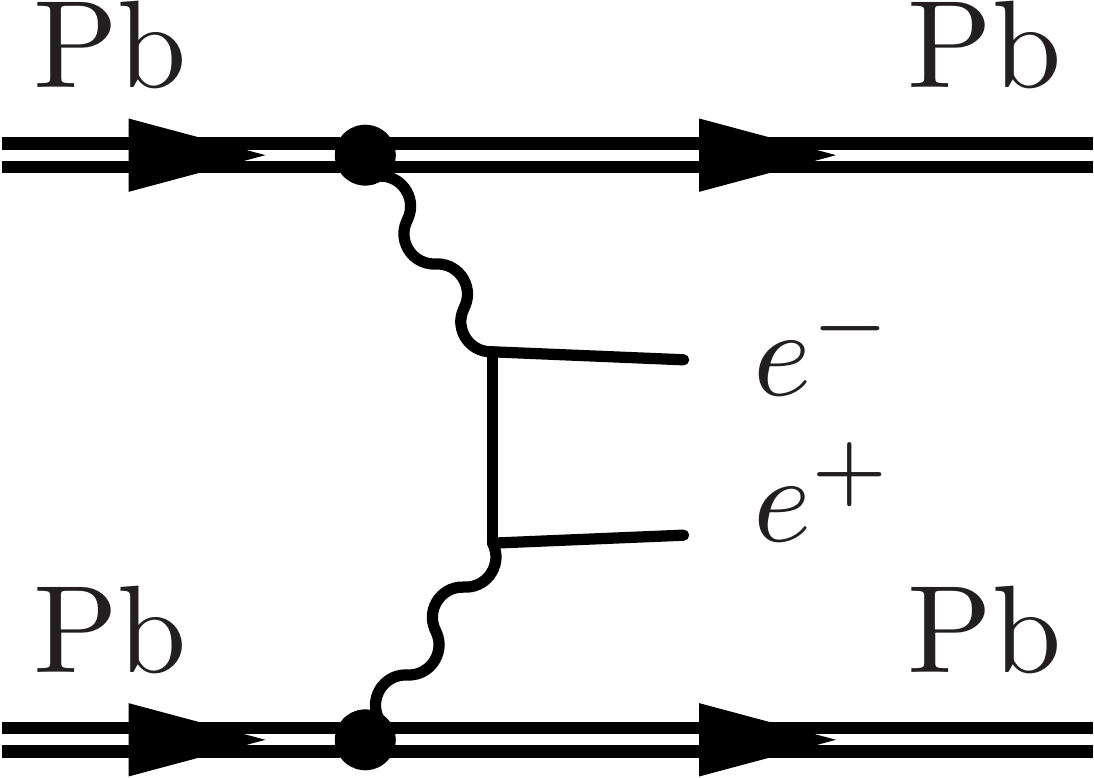}
\caption{The dominant backgrounds to the ALP signal are from light-by-light scattering in ultra-peripheral Pb-Pb collisions, and $e^+ e^-$ production  where both electrons fake a photon.  }
\label{fig:fey_bkg}
\end{figure}

\begin{figure}[t]
\centering
\includegraphics[trim={3.5cm 0.6cm 3cm 0},width=0.28\textwidth]{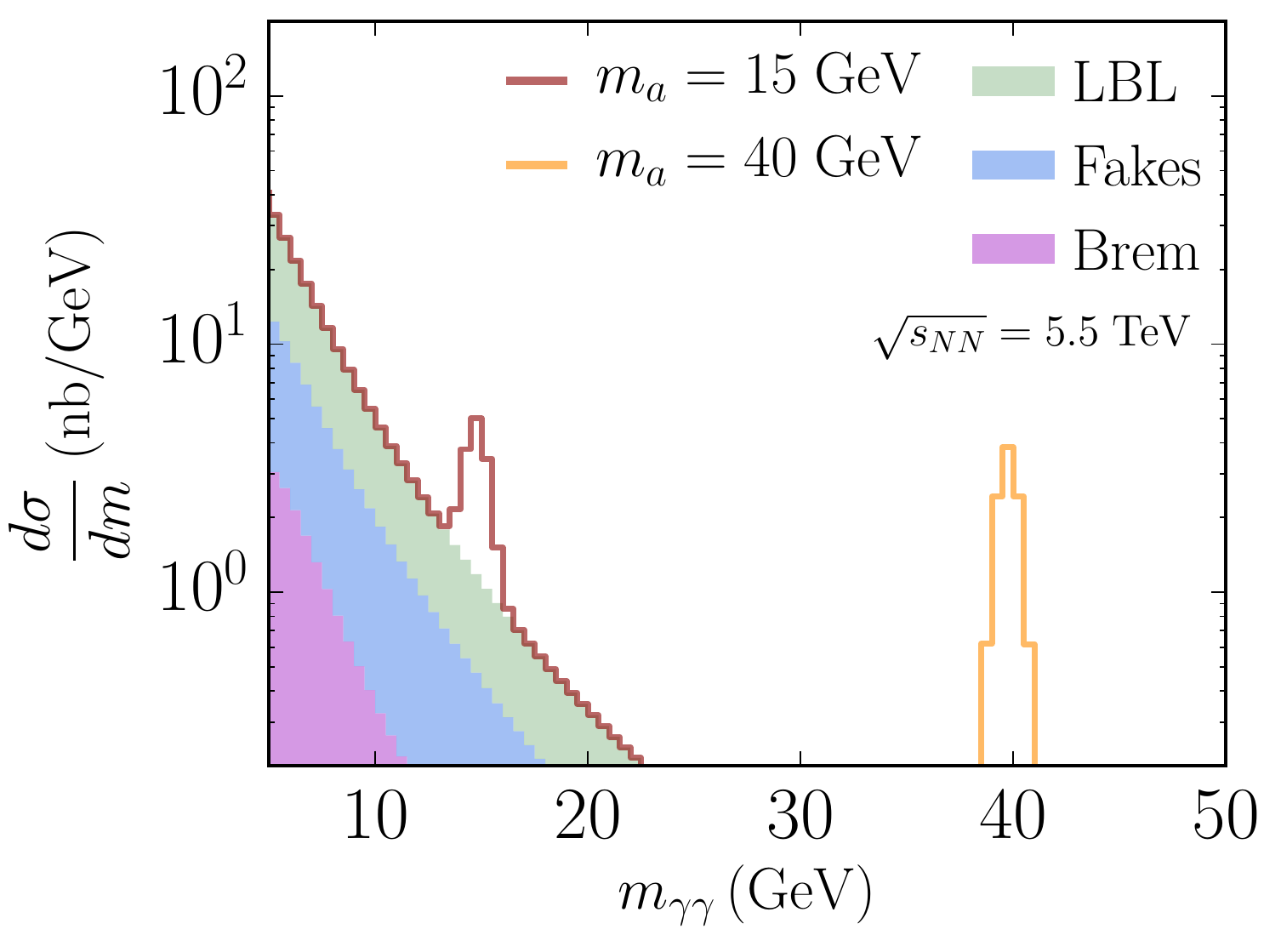}
\caption{
Differential cross section for signal and background. Shown from bottom to top are the stacked background distributions for bremsstrahlung photons from electrons (purple), fake photons from electrons (blue) and light-by-light scattering (LBL) (green). The red (orange) line shows an injected signal with a 5 nb production cross section for $m_a = 15$ GeV and $\Lambda=17$ TeV ($m_a =40$ GeV and $\Lambda=8$ TeV), taking an energy resolution of 0.5 GeV.  
\label{fig:bkg}}
\end{figure}

Due to the $Z^4$ enhancement of the photon flux, the dominant irreducible background comes from light-by-light scattering (LBL), a process which was first calculated for heavy ion collisions in~\cite{d'Enterria:2013yra}. This is shown on the  left-hand side of Fig.~\ref{fig:fey_bkg}. We have computed the rate for LBL in the equivalent photon approximation using the one-loop matrix element for massless fermions~\cite{Bern:2001dg} and find reasonably good agreement with detailed calculations  in~\cite{d'Enterria:2013yra, Klusek-Gawenda:2016euz,Klusek-Gawenda:2016nuo}. Such a background is irreducible but follows a continuum (except for small effects at around the $b\bar{b}$ threshold), as can be seen in Fig.~\ref{fig:bkg}.

Another continuum background where the ions remain intact arises  from exclusive hadronic processes, such as central exclusive production (CEP) of photons. For p-p collisions, this process has been calculated \cite{HarlandLang:2010ep,HarlandLang:2012qz,Harland-Lang:2015cta} and constrained experimentally at 7 TeV with CMS~\cite{Chatrchyan:2012tv}. To the best of our knowledge, no prediction is currently available for the analogous process in heavy-ion collisions. One could make a simplistic estimate of this contribution in Pb-Pb collisions by rescaling the p-p prediction with $\sim A^{2/3}$ (by reason that only the outermost nucleons contribute) which would render this background negligible; however, we note that there is a large theoretical uncertainty in the expected scaling with $A$. Nevertheless, even without an accurate prediction for the rate, this background can be experimentally controlled with the cut of $|\Delta\phi -\pi|<0.04$ applied to the photon pair~\cite{d'Enterria:2013yra}.

 A second hadronic background comes from exclusive production of mesons with substantial branching fractions to photons. We consider exclusive $\pi^0\pi^0$ production as an example process in this category. Using the total rate computed in \cite{KlusekGawenda:2011ib}, we find the fiducial rate after our cuts to be less than 1 nb. In this estimate  we also assumed that two photons for which $\Delta R<0.1$ are resolved as a single photon.

An important reducible background could come from $e^{+}e^{-}$ pair production \cite{Abbas:2013oua}, where both the electron and positron are misidentified as photons. The leading order, fiducial cross section for this process (right-hand panel in Fig.~\ref{fig:fey_bkg}) is as large as 320 $\mu$b, as computed with the \texttt{STARlight} package. 
This large $e^+e^-$ rate implies that it is essential to keep the mis-tag rate sufficiently low. With an estimated $1\%$ mis-tag rate for each electron this process provides a small but non-negligible background, as shown in Fig.~\ref{fig:bkg}.

\begin{figure*}[th!]\centering
	\includegraphics[width=0.47\textwidth]{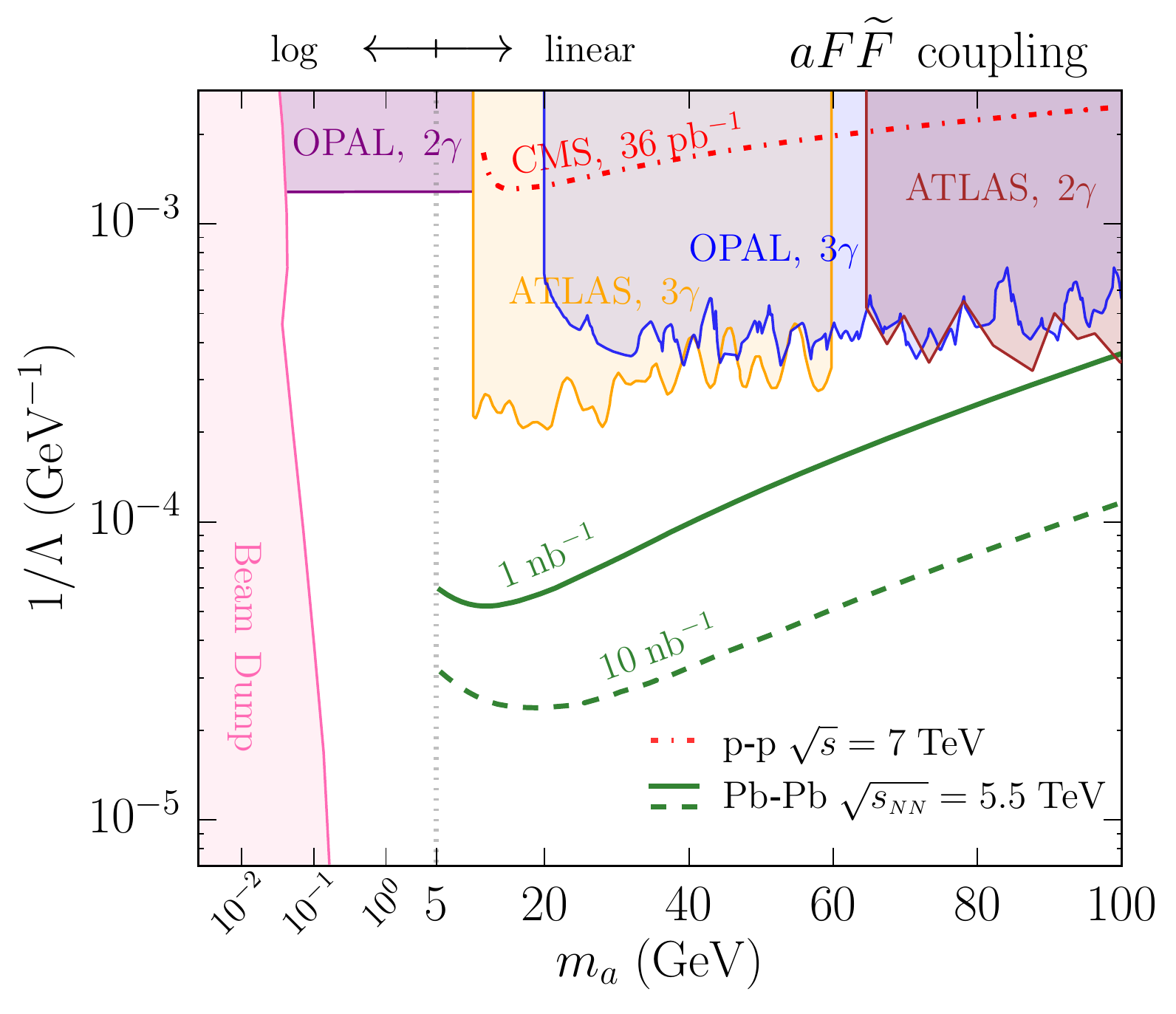}
	\includegraphics[width=0.47\textwidth]{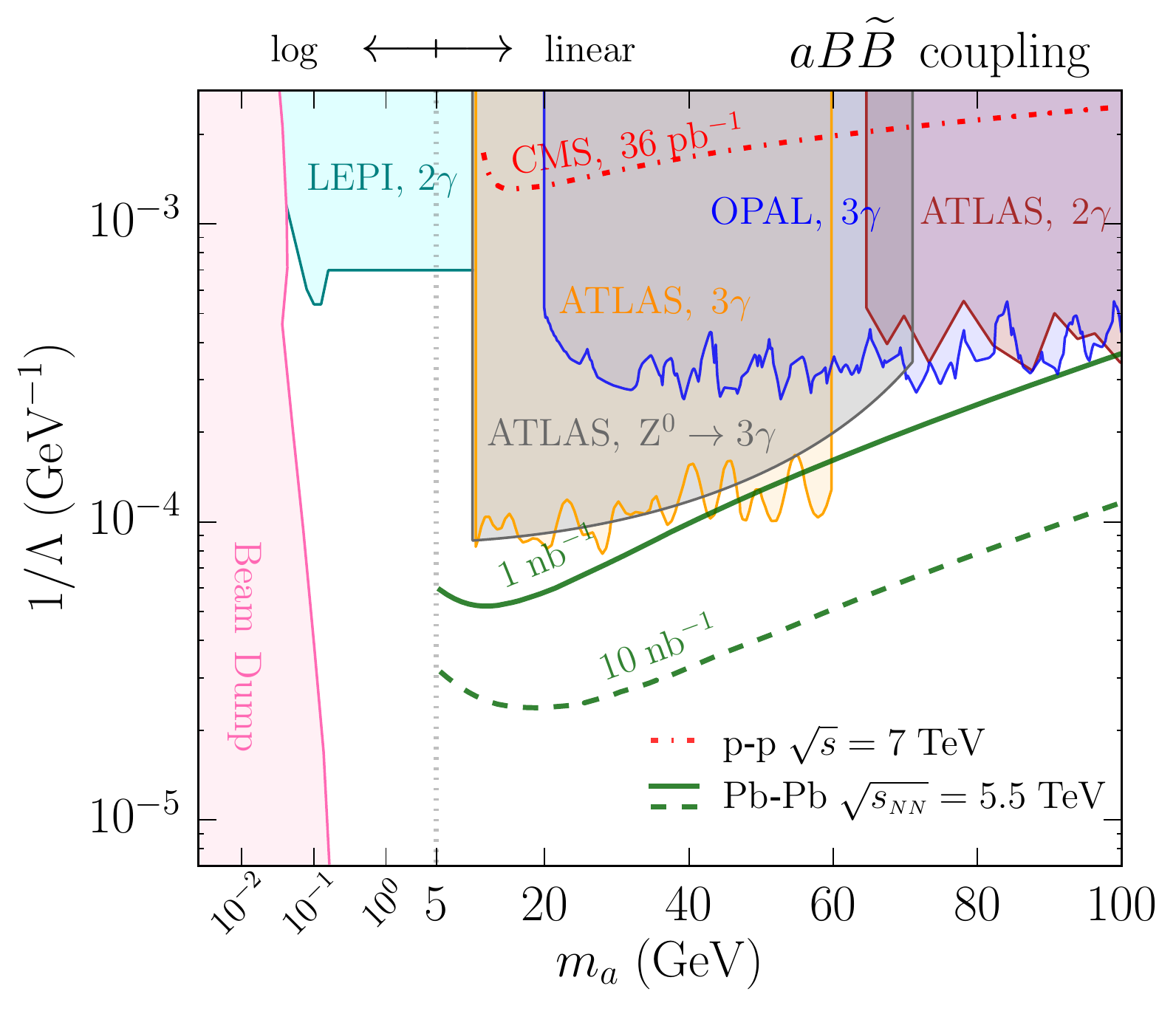}
	\caption{\label{fig:aFF} \emph{Left:} Expected sensitivity to the operator $\frac{1}{4}\frac{1}{\Lambda}a F\tilde F$ in heavy-ion UPCs at the LHC (green solid and dashed curves, for a $\Pb$-$\Pb$ luminosity of 1\,nb$^{-1}$ and 10\,nb$^{-1}$, respectively). Shown for comparison is the limit from 36\,pb$^{-1}$ of exclusive p-p collisions \cite{Chatrchyan:2012tv} (red dot-dash). New and updated exclusion limits from LEPII (OPAL $2 \gamma$,  $3 \gamma$) \cite{Abbiendi:2002je} and from the LHC (ATLAS 2$\gamma$, 3$\gamma$) \cite{Aad:2014ioa,Aad:2015bua} are indicated by the various shaded regions (see text).
\emph{Right}: The analogous results for the operator $\frac{1}{4\cos^2 \theta_W}\frac{1}{\Lambda}a B\tilde B$. The LEPI, $2\gamma$ (teal shaded) region is taken from \cite{Jaeckel:2015jla}.  
 }
\end{figure*}

There could also be a significant number of hard bremsstrahlung photons emitted from the leptons in exclusive $e^{+}e^{-}$ production \cite{Hencken:1999xw} (bremsstrahlung photons from the ions themselves only have $p_T\lesssim 1/R_A \sim 60$ MeV).
Events where the $e^+e^-$ tracks are lost or where both leptons go down the beampipe can then contribute to the background for the $\gamma\gamma$ search. To estimate this contribution, we compute the differential cross section for $\gamma\gamma\rightarrow e^+e^-\gamma\gamma$ for fixed $\sqrt{\hat s}$ with \texttt{MadGraph}~\cite{Alwall:2014hca} and subsequently reweight the cross section with the Pb-Pb photon luminosity function, as discussed in Section~\ref{sec:alps}. We hereby require the $e^+e^-$ to either have high rapidity $|\eta| > 2.5$ or low $p_T < 100$ MeV, while the photons must pass the cuts specified above. Even though the total rate for this process is rather high, we find the fiducial rate to be small, as shown in Fig.~\ref{fig:bkg}.

The relevant exclusive backgrounds and some example signals are all shown in Fig.~\ref{fig:bkg}.
 With an integrated luminosity of 1 $\textrm{nb}^{-1}$ and for $m_{\gamma\gamma}\gtrsim 15$ GeV, we find the expected background to be smaller than 1 event/0.5 GeV. 

\section{Results and discussion}

The ALP parameter space is already substantially constrained by cosmological and astrophysical observations, as well as by a broad range of intensity frontier experiments (see {\it e.g.}~\cite{Essig:2013lka} for a review). In the regime of interest for UPCs (1 GeV $\lesssim m_a\lesssim$ 100 GeV), the existing constraints however come from LEP and LHC \cite{Mimasu:2014nea, Jaeckel:2015jla,Jaeckel:2012yz}.

In Fig.~\ref{fig:aFF} 
we show the expected sensitivity from UPCs, both for the current (1\,nb$^{-1}$) and the high luminosity (10\,nb$^{-1}$) Pb-Pb runs.\footnote{Even though the integrated luminosity is higher, the expected limits from the p-Pb runs are not competitive due to a less favorable $Z^2$ scaling of the rate.  Collisions of lighter ions, {\it e.g.~}Ar-Ar, could be competitive if the integrated luminosity is increased by two to three orders of magnitude compared to Pb-Pb.} For each mass point we computed the expected Poisson limit~\cite{Agashe:2014kda}. In the mass region for which there is background, we assume the entire signal falls into a bin equal to twice the mass resolution. In the remaining, background-free region we set a limit on the total signal rate. 
We also show the analogous limit from the p-p analysis performed by CMS~\cite{Chatrchyan:2012tv}, although we find it is not competitive with other LHC limits. For the $B\tilde B$ operator, the expected limits from heavy-ion collisions are competitive with the other collider limits, whereas for the $F\tilde F$ operator they are significantly stronger.

The existing exclusion limits come from beam dumps \cite{Riordan:1987aw,Dobrich:2015jyk}, LEP and the ({\rm p}-{\rm p}) LHC. We derive LHC limits using a diphoton search at $ m_a > 60$ GeV~\cite{Aad:2014ioa}, and using the ATLAS $3\gamma$ search at lower masses~\cite{Aad:2015bua}.  For the latter search, we  were not able to reliably extract a limit for $m_a\gtrsim 60$ GeV
 with the available public information (see {\it e.g.}~\cite{Mimasu:2014nea, Jaeckel:2015jla,Jaeckel:2012yz} for projected limits). 
For the $B \tilde B$ operator, we also show the limit on the exotic decay $Z^0 \to a \gamma$~\cite{Aad:2015bua}.  

LEP searches also constrain associated production, $e^+e^-\to \gamma a$. We show limits from a resonance search by  OPAL for $m_a > 20$ GeV~\cite{Abbiendi:2002je}. For $50\,$MeV--$8\,$GeV, we derive a new exclusion on the $aF\tilde F$ operator by utilizing data from the OPAL inclusive $2\gamma$ search~\cite{Abbiendi:2002je}. This improves on previous limits~\cite{Jaeckel:2015jla}, which were derived using LEPI data. The analogous LEPI limits from \cite{Jaeckel:2015jla} are shown for the $aB\tilde B$ operator. Appendix~\ref{appendix} gives more details on the LEP and LHC limits summarized above.


Below $m_a\lesssim 5$ GeV the signal in Fig.~\ref{fig:fey_alp} falls below the trigger thresholds, and it is an interesting puzzle as to how the reach can be extended to this regime. 
To further probe this region with Pb-Pb collisions, we considered: {\it i)} an off-shell $a$ would provide a new contribution to light-by-light scattering;  {\it ii)} associated production, for example with electrons $\gaga\to a\,e^+e^-$; and, {\it iii)} ALP pair-production  $\gaga\to aa$. Unfortunately, these signal cross sections do not provide enough sensitivity compared to existing constraints: for $\Lambda=1\,$TeV we find 0.004\,nb, 0.2\,nb and 0.01\,nb, respectively.

In summary, we have found that heavy-ion collisions at the LHC can provide the best limits on ALP-photon couplings for $5\,\mathrm{GeV}<m_a<100\,\mathrm{GeV}$. The very large photon flux and extremely clean event environment in heavy-ion UPCs provides a rather unique opportunity to search for BSM physics.

\section*{Acknowledgements}

We are grateful to Bob Cahn, Lucian Harland-Lang, Yonit Hochberg, Joerg Jaeckel, Spencer Klein, Hitoshi Murayama, Michele Papucci, Dean R\"obinson, Sevil Salur, and Daniel Tapia Takaki for useful discussions. 
SK and TL thank the participants of the 3rd NPKI workshop in Seoul for useful discussions. We further thank Spencer Klein and Daniel Tapia Takaki for valuable comments on the manuscript.
The authors are supported by DOE contract DE-AC02-05CH11231. TL is also supported by NSF grant PHY-1316783. 

\appendix

\section{Details on LEP and LHC limits\label{appendix}}

{\bf{LEP.}}  The LEPI limits shown in the right panel of Fig.~\ref{fig:aFF} are taken from Ref.~\cite{Jaeckel:2015jla}, which used an inclusive $e^+ e^- \to 2\gamma$ search on the $Z$-pole to set limits on the process $ e^+ e^- \to Z \to a \gamma$.

We extract limits from LEPII using the OPAL analysis $e^+e^- \to 2 \gamma, 3\gamma$~\cite{Abbiendi:2002je}. For $m_a > 20$ GeV, we apply the limits on the cross section for $e^+e^- \to a \gamma, a \to 2\gamma$ given in Fig.~9 of Ref.~\cite{Abbiendi:2002je},  obtained using the data sample with three photon candidates. In the mass range $50\,$MeV--$8\,$GeV, we instead use the inclusive $e^+ e^- \to 2 \gamma$ signal region. For such ALP masses, the photons from the ALP decay are collimated but  no explicit photon isolation is required by the analysis of~\cite{Abbiendi:2002je}.  We derive new limits by generating events for $e^+e^- \to a \gamma, a \to 2\gamma$ using {\tt MadGraph}, and then applying the selection criteria of~\cite{Abbiendi:2002je} on photon energy, angle, and acoplanarity (finding an efficiency $\sim$0.9). Then, using the observed and expected background quoted in Tab.~5 of~\cite{Abbiendi:2002je}, and assuming Poisson statistics, we set a 95\% confidence limit bound on the signal cross section.

{\bf{LHC.}} For $m_a>60$ GeV, we calculate the fiducial cross section of $p p\rightarrow a\rightarrow \gamma\gamma$ and compare with the constraints in Ref.~\cite{Aad:2014ioa}.  Associated production of $a \gamma$ is also constrained by multi-photon searches at the LHC. The ATLAS analysis in Ref.~\cite{Aad:2015bua} considers rare Higgs and $Z$ decays to three or more photons at the $\sqrt{s} = 8 $ TeV run of the LHC.  While the models considered are somewhat different from ours, the search for Higgs decay to $aa$ has a similar signal region. This analysis requires three photons and places limits on a resonance in the invariant mass of second and third photon. We have approximately recast the published limits on  the Higgs decay $ h \to a a$  by rescaling the acceptances for the photon cuts from the Higgs model to the case of ALP associated production.  This is shown as the yellow region labeled ATLAS $3\gamma$ in our figures. With the available public information, it is however not possible to fully reconstruct the analysis for the ALP case, in particular due to the different kinematics and combinatorics in the final state. We expect that this limit can therefore be made more robust with a dedicated study by the collaboration.

For the $B \tilde B$ operator, we repeat the analysis described above, including the contribution from $Z^0$ exchange. In addition $Z^0 \to a \gamma$ is also possible as an exotic $Z^0$ decay. 
We obtain a limit by comparing the branching ratio with the constraint on $Z^0\rightarrow 3\gamma$~\cite{Aad:2015bua}, shown as the $Z^0 \to 3 \gamma$ region in the right-hand panel of Fig~\ref{fig:aFF}. We have cut this off at $m_a \approx 70$ GeV due to the requirement of $p_T^\gamma > 17$ GeV.  

\bibliographystyle{utphys}
\bibliography{alps}

\end{document}